\documentclass{article}
\usepackage{amsmath,amssymb}
\usepackage{dsfont}
\usepackage{epsfig}
\usepackage{epsf} 
\usepackage{graphicx}
\usepackage{slashed}
\usepackage{color}
\newcommand{\beq}{\begin{equation}}
\newcommand{\eeq}{\end{equation}}
\newcommand{\bea}{\begin{eqnarray}}
\newcommand{\eea}{\end{eqnarray}}
\newcommand{\wt}{\widetilde}

\newcommand{\wh}{\widehat}

\begin{document}

\begin{center}

~\vspace{0.2cm}

{\bf STRONG COUPLING FROM THE TAU HADRONIC WIDTH  \\BY NON-POWER  QCD PERTURBATION THEORY}\footnote{Contribution to the proceedings of the workshop "Determination of the Fundamental Parameters of QCD", Nanyang Technological University, Singapore, 18-22 March 2013, to be published in Mod. Phys. Lett. A.}
\\

\vspace{0.8cm}
 I. CAPRINI\\
\vspace{0.2cm}

Horia Hulubei National Institute for Physics and Nuclear Engineering,\\
P.O.B. MG-6, 077125 Bucharest-Magurele, Romania

\end{center}
\vspace{0.2cm}

\begin{abstract}
 Starting from the divergent character of the perturbative expansions in QCD and using the technique of series acceleration by the conformal mappings  of the Borel plane, I define  a novel, non-power perturbative expansion for the Adler function, which simultaneously  implements renormalization-group summation and has a tamed large-order behaviour.
The new expansion functions, which replace the standard powers of the coupling, are singular at the origin of the coupling plane  and have divergent perturbative  expansions,  resembling the expanded function  itself.  Confronting the new perturbative expansions with the standard ones on specific models investigated recently in the literature, I show that they approximate
in an impressive way the exact Adler function and  the spectral function moments.  Applied to the $\tau$ hadronic width,  the contour-improved and the renormalization-group summed non-power expansions in the ${\overline{\rm MS}}$ scheme lead to the prediction $\alpha_s(M_\tau^2)= 0.3192~^{+ 0.0167}_{-0.0126}$, which translates to $\alpha_s(M_Z^2)= 0.1184~^{+0.0020}_{-0.0016}$. 
\end{abstract}

\section{Introduction}	The hadronic decays of the $\tau$ lepton allow  one of the most precise determinations of the strong coupling and provide also a beautiful experimental test of the predicted QCD running \cite{RPP,Pich2013}.  However, this process involves the strong coupling at a relatively low scale, where the theoretical ambiguities inherent to perturbative QCD  are expected to be large. This is why the great accuracy claimed in some of the earlier determinations of $\alpha_s(M_\tau^2)$ was questioned by some authors \cite{Altarelli2013}.

The situation improved with the recent investigations \cite{Davier2008}-\cite{BeBoJa},  motivated  by the calculation of the Adler function in massless QCD to four loops \cite{BCK08}.  In these works (and many other papers, not quoted here due to lack of space), various sources of ambiguity of perturbative QCD were examined in detail and a more realistic estimate of the error was given. In this talk I will consider the determination of the strong coupling from the $\tau$ hadronic width, $R_\tau= \Gamma[\tau^-\to \nu_\tau+{\rm hadrons}\, (S=0)]/  \Gamma[\tau^-\to\nu_\tau e^-\bar\nu_e]$. Restricting the analysis to the dominant Cabibbo allowed decay width related to the $V+A$ correlator, analyticity and unitarity lead to the theoretical expression 
 \beq R_{\tau, V+A} \,=\, N_c\,S_{\rm EW}\,|V_{ud}|^2\,\biggl[\,
1 + \delta^{(0)}  + \sum\limits_{D\geq 2} 
\delta_{ud}^{(D)} \,+ \delta_{\rm EW} \biggr], \eeq
 where $N_c$ is the number of quark colours,
$S_{\rm EW}$  and $\delta_{\rm EW}$  are electroweak corrections,  $\delta_{ud}^{(D)}$ denote quark mass
and higher $D$-dimensional nonperturbative  corrections, and  $\delta^{(0)}$ is the $D=0$ perturbative contribution, written as 
 the contour integral
\beq \label{eq:delta0}\delta^{(0)}=\frac{1}{2 \pi i}\!\! \oint\limits_{|s|=M_\tau^2}\!\! \frac{d s}{s}
\left(1- \frac{s}{M_\tau^2}\right)^3 \left(1+\frac{s}{M_\tau^2}\right) \widehat{D}_{pert}(s), \eeq
in terms of the perturbative expansion of the reduced Adler function $ \wh D(s)= -s\,{\rm d}/{\rm d}s\,\Pi^{(0+1)}(s)-1$.

The kinematic weight in the integral (\ref{eq:delta0}) suppresses the nonperturbative power corrections (PC) and the effects of duality-violation (DV),  {\em i.e.} the breakdown of the operator product expansion (OPE) near the timelike axis in the complex energy plane. This makes the hadronic width a good observable for extracting the strong coupling. 
However there are still two important theoretical uncertainties: first, the truncated series  $\widehat{D}_{pert}(s)$ depends on the prescription chosen for implementing the renormalization-group invariance (RGI). Second,  the coefficients of the perturbative series display a factorial growth, so the series has a vanishing radius of convergence.  
The two uncertainties are actually related and  affect the precision of the predictions at the $M_\tau$ scale, where the coupling is rather large.  In particular, it was noticed that the inclusion of additional terms  in the expansion did not reduce, but on the contrary increased the dependence on the renormalization-group prescription. This raises the question of whether a  new type of perturbative expansions, which simultaneously impose RGI and the theoretical knowledge on the large-order behaviour of the series, can be found. The present talk addresses precisely this question.
 After a brief review of the prescriptions used for RG implementation and the large order behaviour of the series, I introduce a new type of perturbative expansions in QCD, which replace the powers of the coupling by more complicated non-power functions,  and use them for the extraction of $\alpha_s$ from the $\tau$ hadronic width.
 
\section{FOPT, CIPT and RGSPT}\label{sec:pt}
 The standard perturbative expansion of the Adler function in a definite renormalization scheme is written as
\beq \label{eq:fopt}
\widehat{D}_{\rm FOPT}(s) = \sum\limits_{n\ge 1} (a_s(\mu^2))^n\,[c_{n,1}+
\sum\limits_{k=2}^{n} k\, c_{n,k}\, (\ln (-s/\mu^2))^{k-1}],~ a_s(\mu^2)=\frac{\alpha_s(\mu^2)}{\pi}, \eeq
 and is  denoted usually as "fixed-order perturbation theory" (FOPT) \cite{BeJa}.  In (\ref{eq:fopt}) the renormalization scale $\mu^2$ is  chosen near $M_\tau^2$,  the coefficients $c_{n,1}$  are calculated from Feynman diagrams,  and  $c_{n,k}$ for $2 \leq k \leq n $ depend on $c_{n,1}$ and  the perturbative coefficients $\beta_k$  of the RG $\beta$-function, which are known at present to four loops.  In the  ${\overline{\rm MS}}$ scheme for $n_f=3$ flavours  the coefficients $c_{n,1}$  calculated up to know (cf. \cite{BCK08} and references therein) are:
\beq\label{eq:cn1} c_{1,1}=1,~ c_{2,1}= 1.64,~ c_{3,1}=6.371,~ c_{4,1}=49.079.\eeq
By taking $\mu^2=-s$ in (\ref{eq:fopt}), one obtains the  "renormalization-group-improved", denoted also as  ``contour-improved perturbation theory" (CIPT) \cite{Pich_Manchester}. It is written as 
\beq \label{eq:cipt}\widehat{D}_{\rm CIPT}(s)= \sum\limits_{n\ge 1} c_{n,1} \,(a_s(-s))^n, ~~~~~ \eeq
where the running coupling  $a_s(-s)$  is determined by solving the renormalization-group equation $s\,{\rm d} a_s(-s)/{\rm d}s = \beta (a_s)$ numerically in an iterative way along the circle, 
starting with the input value $a_s(M_\tau^2)$ at $s=-M_\tau^2 $.

 The properties of the above expansions, in particular their convergence  and the behaviour in the complex energy plane, were analysed in detail in several recent papers \cite{Davier2008, BeJa,Pich_Manchester, Beneke_Muenchen,BeBoJa}.

I mention also another prescription, proposed in \cite{Ahmady} for timelike observables and applied in  \cite{Abbas:2012,Abbas:2013} for Adler function in the complex plane. It is written as: 
\beq \label{eq:rgspt}\wh D_{\rm RGSPT}(s) = \sum_{n\ge 1} \,(\wt a_s(-s))^n [c_{n,1}+ \sum_{j=1}^{n-1} c_{j,1} d_{n,j}(y)], \eeq
where $\wt a_s(-s)=a_s(\mu^2)/(1+\beta_0 a_s(\mu^2) \ln(-s/\mu^2))$  is the solution of the RG equation  to one loop and 
 $d_{n,j}(y)$ are  calculable functions depending on the variable $y\equiv 1+\beta_0 a_s(\mu^2) \ln(-s/\mu^2)$ and the coefficients  $\beta_j$, which vanish for $y=1$ or in the limit $\beta_j=0$, $j\ge 1$. I refer to this as "renormalization-group summed perturbation theory" (RGSPT).  As an effective series in powers of the one-loop running coupling, with coefficients that depend on the nonleading $\beta_j$  and  the coupling at a fixed scale, it appears to be  "in-between" FOPT and CIPT. For more details see  \cite{Abbas:2012,Abbas:2013,Abbas:proc}. 
\section{Large order behaviour and the Borel transform}
From special classes of Feynman diagrams  it is known  that the perturbative coefficients $c_{n,1}$ display a factorial increase, $c_{n,1}\sim n!$, so the perturbative expansions written above are divergent series. They are often interpreted as asymptotic series  \cite{Muell,Beneke}. This property follows also indirectly from the  arguments given in \cite{tHooft}, which infer that the expanded amplitude, viewed as a function of the coupling $\alpha_s$, is singular at $\alpha_s=0$.

 The large order behaviour of the series (\ref{eq:cipt}) is encoded in the properties of the Borel transform $B(u)$, defined by the expansion 
\beq\label{eq:B}
B(u)= \sum_{n=0}^\infty c_{n+1,1}\, \frac{u^n}{\beta_0^n \,n!}.
\eeq
The original function $\wh D_{\rm CIPT}(s)$  is recovered from $B(u)$ by a Laplace-Borel integral. Actually, in the present case $ B(u)$ is known to have singularities on the real axis of the $u$-plane, more precisely along the lines $u\ge 2$ and $u\le -1$ \cite{Beneke}, so the integral requires a prescription. I adopt here  the principal value ({\rm PV}) prescription \cite{Muell,Beneke,BeJa}  
\beq\label{eq:LaplaceCI}
 \wh D_{\rm CIPT}(s)=\frac{1}{\beta_0}\,{\rm PV} \,\int\limits_0^\infty  
\exp{\left(\frac{-u}{\beta_0 a_s(-s)}\right)
} \, B(u)\, {\rm d} u\,,
\eeq
which is preferred from the point of view of momentum-plane analyticity \cite{CaNe}.

 Similarly, one defines the Borel transforms $B_{\rm FO}(u,s)$ and $B_{\rm RGS}(u, y)$  of the FOPT and RGSPT expansions, (\ref{eq:fopt}) and (\ref{eq:rgspt}) respectively, which can be written as \cite{Abbas:2013}: 
\bea\label{BFO}
&&B_{\rm FO}(u,s)=B(u)+ \sum_{n=0}^\infty  \frac{u^n}{\beta_0^n \,n!} \sum\limits_{k=2}^{n+1} k\, c_{n+1,k}\,\left(\ln\frac{-s}{M_\tau^2}\right)^{k-1},\\
&&B_{\rm RGS}(u, y)=B(u) + \sum_{n=0}^\infty\frac{ u^n} {\beta_0^n \,n!} \sum_{j=1}^{n} c_{j,1} d_{n+1,j}(y).\nonumber
\eea
The functions $\wh D_{\rm FOPT}(s)$ and $\wh D_{\rm RGS}(s)$ are recovered from their Borel transforms  by Laplace-Borel integrals similar to (\ref{eq:LaplaceCI}). It is important to recall that not only the location, but also the nature of the leading singularities of $B(u)$ is known. Namely, near these points $B(u)$ behaves as 
\beq\label{eq:Bthr}
B(u) \sim (1+u)^{-\gamma_{1}},  \quad B(u)  \sim (1-u/2)^{-\gamma_{2}}, \eeq
where  $\gamma_1 = 1.21$ and  $\gamma_2 = 2.58$  \cite{Muell,Beneke,BeJa}. As argued in \cite{Muell,Abbas:2013}, the Borel transforms $B_{\rm FO}(u,s)$ and $B_{\rm RGS}(u, y)$ have the same leading singularities in the $u$-plane.

\section{Need for a new perturbative expansion}\label{sec:need}
As mentioned above, the function $\wh D_{pert}$ is expected to be singular at the expansion point $a_s=0$, while the integer powers of $a_s$ are all holomorphic. Thus, no finite-order standard perturbative approximant can share the singularity
with the expanded function at zero coupling: singularities can emerge only from the infinite 
series as a whole, which, unfortunately, is not defined, since the perturbation series is divergent.

As discussed in \cite{CaFi2011}, a perturbation series would be more instructive if the
finite-order approximants could retain some information about the 
known singularities of the expanded function.  Such approximants would tell us more about the function
also from the numerical point of view. In the next section I will explain how such an expansion can be defined, using the idea of analytic continuation and series acceleration by conformal mappings.
\section{Non-power perturbative expansions}\label{sec:nppt}
The starting point in the derivation is the remark that the expansion (\ref{eq:B}) converges only in the disk $|u|<1$, whose boundary passes through the  singularity of $B(u)$ nearest to $u=0$. However, the function $B(u)$ is holomorphic in a larger domain, assumed in general to be the whole complex $u$-plane cut along the lines $u\ge 2$ and $u\le -1$. It would be useful to insert in (\ref{eq:LaplaceCI}) an expansion of $B(u)$  that is convergent also outside the disk $|u|<1$. Such an expansion is easily obtained: since the disk is the natural domain of convergence  for power series, it suffices to expand the function in powers of variables that map a larger part of its holomorphy domain   onto a disk.  Intuitively, one expects that a larger  domain of convergence is related also to a better convergence rate. This expectation was confirmed by a mathematical result given in \cite{CiFi},  which proves the existence of an "optimal conformal mapping" (OCM) for series expansions. It is the variable that maps the entire holomorphy domain of the expanded function onto a disk, and has the remarkable (less-known) property that by expanding in powers of this variable one obtains the series with the fastest large-order convergence rate at all points inside the holomorphy domain. More detailed arguments are given in two lemmas  formulated and proved in \cite{CaFi_Manchester,CaFi2011}. For the Adler function in QCD, the optimal mapping $\wt w(u)$ and the corresponding perturbative expansion were proposed and investigated in \cite{CaFi1998,CaFi2001}. 

An additional improvement is obtained by exploiting the known
 behaviour (\ref{eq:Bthr}) of $B(u)$ near the first singularities. 
If one multiplies $B(u)$ with a suitable factor  $S(u)$, which "softens" the dominant singularities, the expansions will have a more rapid convergence. In fact,  the presence of mild branch-points, where the function vanishes instead of becoming infinite, is expected to become visible only at large orders in  the power expansions of the function. Hence, one can  expand the product $S(u) B(u)$ in powers of conformal mappings that account  for the nonleading, {\em i.e} the more distant, singularities, and contain a residual "mild" cut inside the convergence disks. In view of these remarks, it was useful to consider the  general class of conformal mappings  \cite{CaFi2011}:
\beq \wt w_{jk}(u)=\frac{\sqrt{1+u/j}-\sqrt{1-u/k}}{\sqrt{1+u/j}+\sqrt{1-u/k}},\eeq
where $j, k$ are positive integers satisfying  $j\ge 1$ and $k \ge 2$.  The function $\wt w_{jk}(u)$ maps the $u$-plane cut along $ u\le -j$ and $u\ge k$ onto the disk $|w_{jk}|<1$ in the plane $w_{jk}\equiv \wt w_{jk}(u)$. The optimal mapping defined above is $\wt w(u)\equiv \wt w_{12}(u)$, for which the entire holomorphy domain of the Borel transform is mapped onto the interior of the unit circle in the plane $w_{12}$. 

Using the above ideas, the following expansion was introduced in \cite{CaFi_Manchester,CaFi2011}
\beq\label{eq:Bw}
 S_{jk}(u) B(u) =\sum_n c_{n, {\rm CI}}^{(jk)}\,  (\wt w_{jk}(u))^n,
\eeq
with the "softening factors"  chosen as
\beq  S_{jk}(u)=(1-\wt w_{jk}(u)/\wt w_{jk}(-1))^{\gamma'_1} \left(1-\wt w_{jk}(u)/\wt w_{jk}(2)\right)^{\gamma'_2} \eeq
where $ \gamma_j'$, $j=1,2$, are suitable exponents defined such as to preserve the behaviour (\ref{eq:Bthr}) of $B(u)$. The expansions 
(\ref{eq:Bw}) converge in a larger domain compared with the original series (\ref{eq:B}), and according to the lemmas proven in \cite{CaFi2011},  have a better convergence rate, in particular at points $u$ close to the origin, which are dominant in the  Laplace integral (\ref{eq:LaplaceCI}). 
The use of several conformal mappings and different softening factors  reduces the bias related to the implementation of the threshold behaviour (\ref{eq:Bthr}), which is not unique \cite{CaFi2011}.  In practice, as discussed in \cite{CaFi2011}, useful mappings are $\wt w_{12}(u)$ (denoted above as OCM), $\wt w_{13}(u)$,  $\wt w_{1\infty}(u)$ and $\wt w_{23}(u)$. 

 From (\ref{eq:LaplaceCI}) and (\ref{eq:Bw})  one
obtains the "contour-improved non-power perturbation theory" (CINPPT) \cite{CaFi2011}
\beq\label{eq:cinppt} \wh D_{\rm CINPPT} (s)= \sum\limits_{n \ge 0} c_{n, {\rm CI}}^{(jk)} \, {\cal W}^{jk}_{n, {\rm CINPPT}}(s),\eeq 
where the expansion functions are defined as
\beq\label{eq:Wnpci}  {\cal W}^{jk}_{n, {\rm CINPPT}}(s)=\frac{1}{\beta_0} {\rm PV} \int\limits_0^\infty\!{\rm e}^{-u/(\beta_0 a_s(-s))} \,  \frac{(\wt w_{jk}(u))^n}{S_{jk}(u)}\,{\rm d} u. \eeq
Similarly, the "fixed-order non-power perturbation theory" (FONPPT) and the "renormalization-group summed non-power perturbation theory" (RGSNPPT) are defined as \cite{CaFi2011,Abbas:2013}
\beq\label{eq:fonppt} \wh D_{\rm FONPPT}  (s) = \sum\limits_{n\ge 0} c_{n, {\rm FO}}^{(jk)}(s) \,{\cal W}^{jk}_{n, {\rm FONPPT}},\eeq
\beq \label{eq:rgsnppt} \wh D_{\rm RGSNPPT}  (s) = \sum\limits_{n\ge 0} c_{n, {\rm RGS}}^{(jk)}(y) \, {\cal W}^{jk}_{n, {\rm RGSNPPT}}(s), \eeq
where the expansion functions ${\cal W}^{jk}_{n, {\rm FONPPT}}$ and ${\cal W}^{jk}_{n, {\rm RGSNPPT}}(s)$ are obtained from (\ref{eq:Wnpci}) by replacing $a_s(-s)$ in the exponent through  $a_s(M_\tau^2)$ and $\tilde a_s(-s)$, respectively.

The properties of the expansions (\ref{eq:cinppt})-(\ref{eq:rgsnppt}) have been discussed in \cite{CaFi1998,CaFi2001,CaFi2011}. A remarkable feature is that
 the expansion functions resemble the expanded function, being singular at   $a_s=0$ and having divergent series in powers of $a_s$. Therefore, the divergent pattern of the expansion of the QCD correlators in terms of these
new functions is expected to be tamed.

\section{Confronting the new expansions with the standard ones}\label{sec:num}
It is instructive to compare the new expansions with the standard ones at various perturbative orders.  A class of models  discussed recently in the literature \cite{BeJa,CaFi2011,BeBoJa}, which describe the Adler function in terms of a small number of dominant renormalons, offers a good framework for this exercise. Due to space limitations, I consider only the so-called "reference model" proposed in \cite{BeJa,BeBoJa}.

\begin{figure}
\begin{center}\includegraphics[width=6.cm]{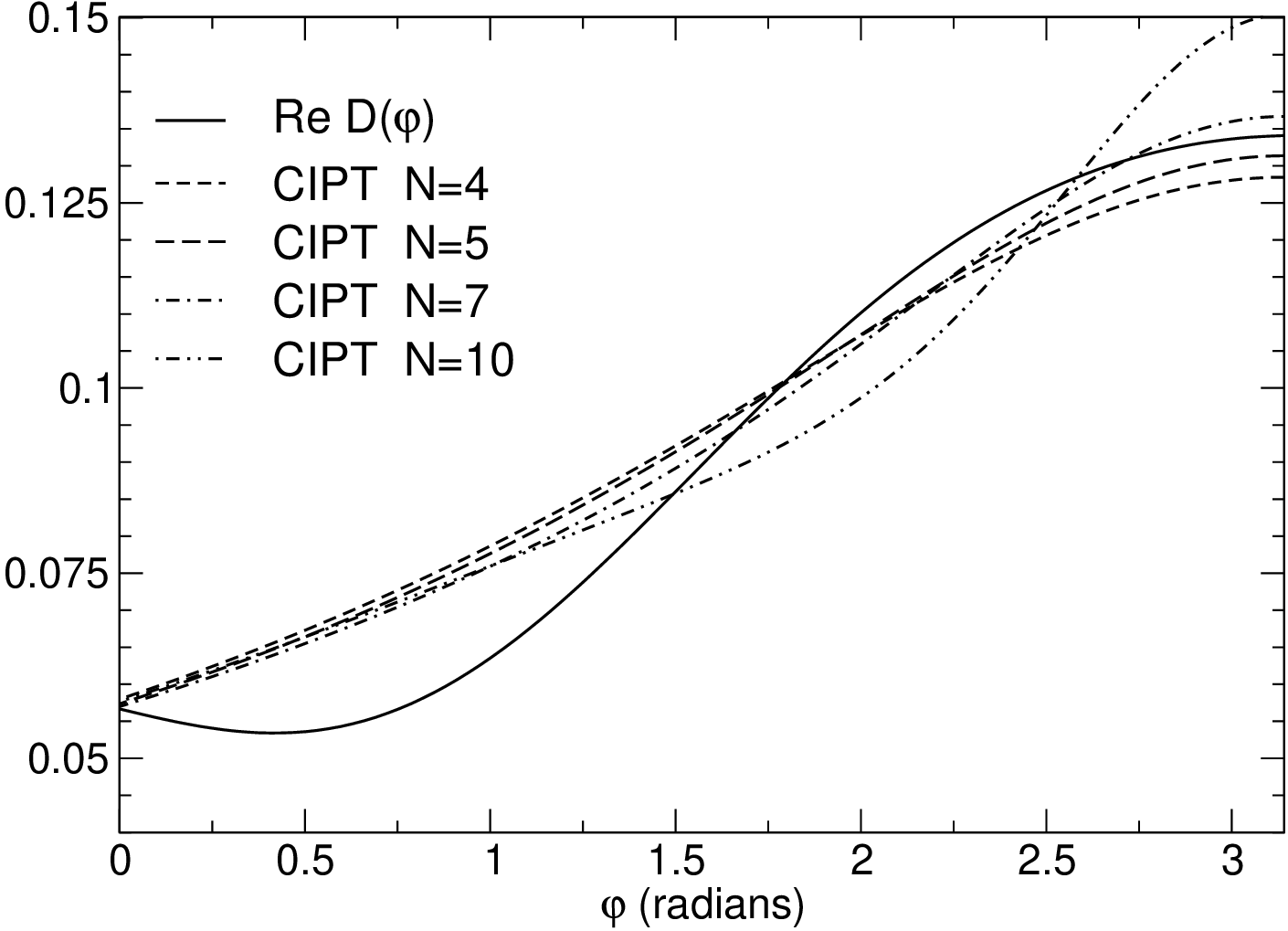}\hspace{1.5cm}
\includegraphics[width=6.cm]{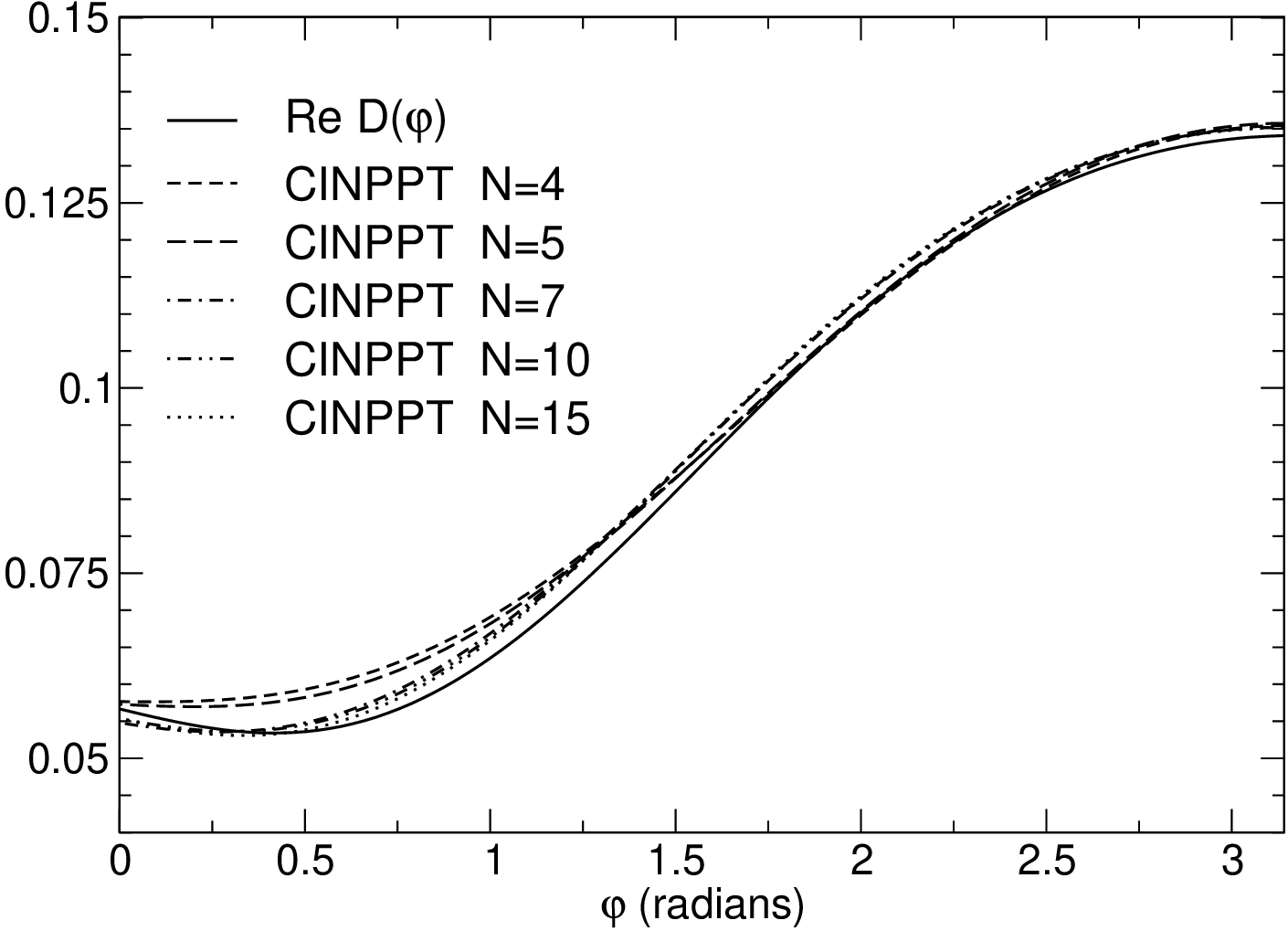}\caption{Real part of the Adler function of the reference model \cite{BeJa,BeBoJa} along the circle $|s|=M_\tau^2$, calculated with the standard CIPT (left) and the optimal CINPPT (right).  }\label{fig1}
\end{center}\end{figure}

\begin{figure}
\begin{center}\includegraphics[width=6.cm]{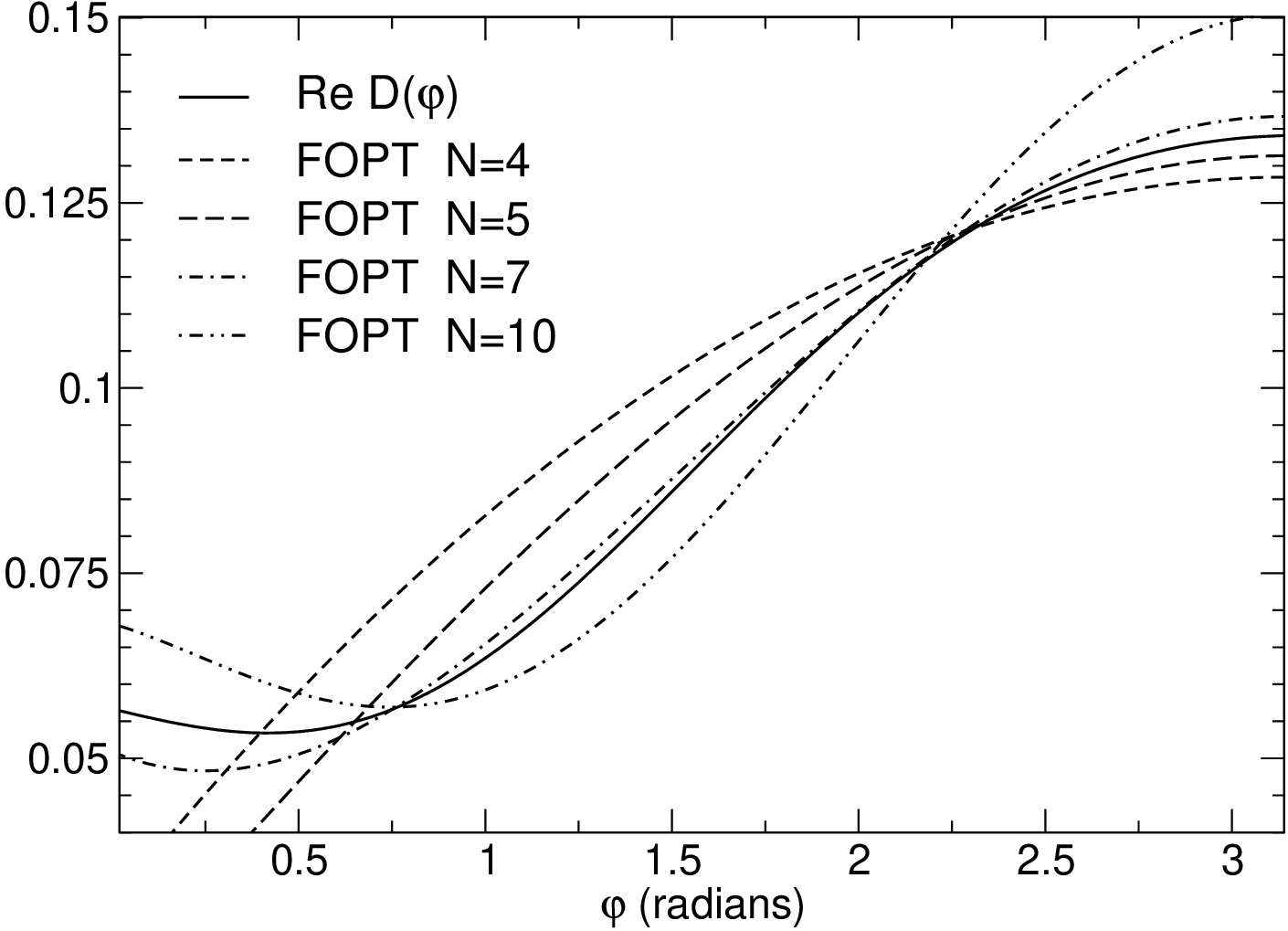}\hspace{1.5cm}
\includegraphics[width=6.cm]{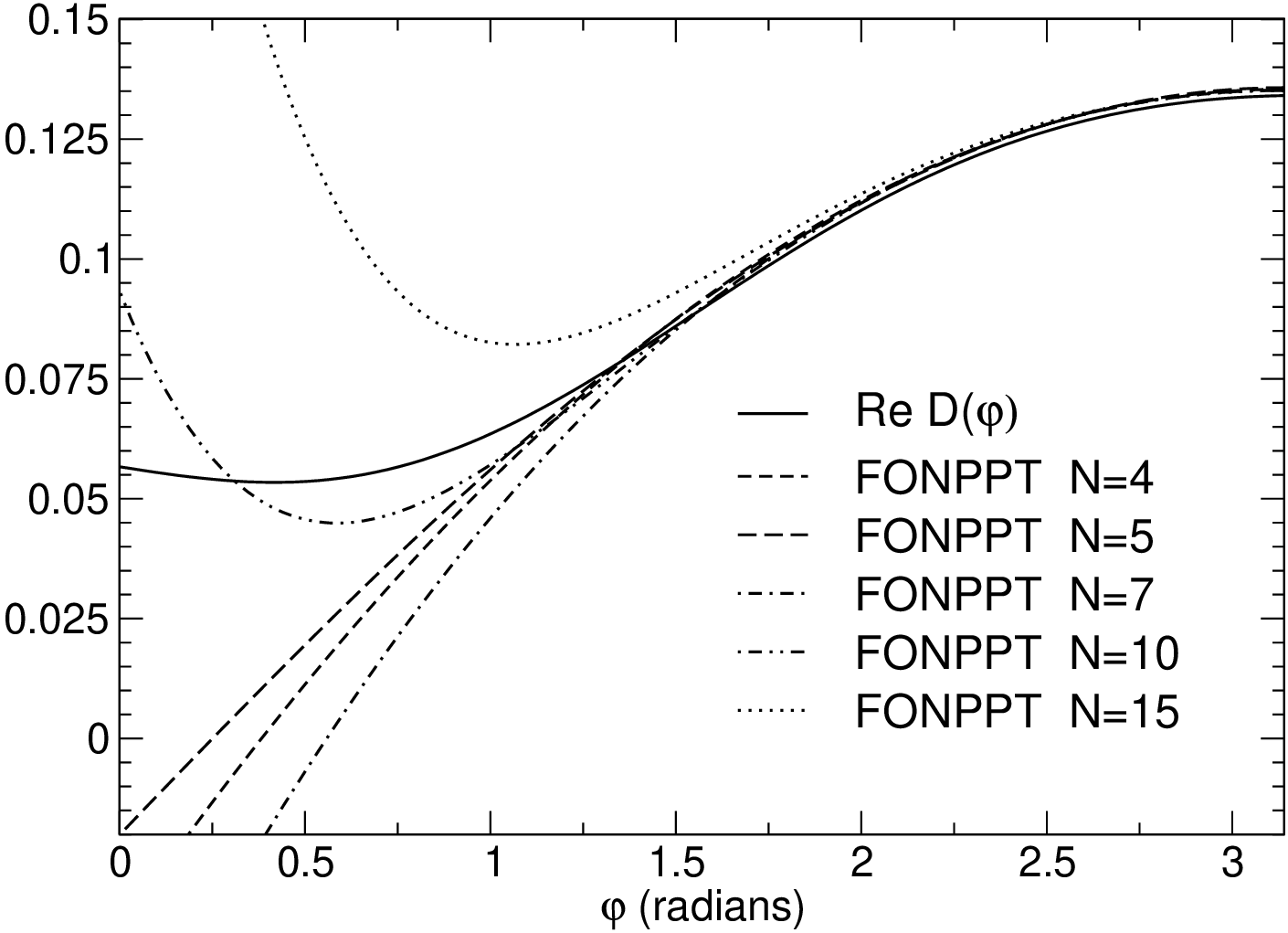}\caption{Real part  of the Adler function of the reference model \cite{BeJa,BeBoJa} along the circle $|s|=M_\tau^2$, calculated with the standard FOPT (left) and the optimal  FONPPT (right).}\label{fig2}\end{center}\end{figure}

 Figs. \ref{fig1} and \ref{fig2}  illustrate the approximation of the real part of the Adler function of the model with the standard CIPT and FOPT and the corresponding non-power expansions  (\ref{eq:cinppt}) and (\ref{eq:fonppt}) truncated at a perturbative  order $N$.  The  optimal conformal mapping (OCM) $\wt w_{12}$ and the input value $\alpha_s(M_\tau^2)=0.34$ were used in this exercise.

One can see the very good approximation  along the whole circle $|s|=M_\tau^2$ provided by CINPPT up to high perturbative orders $N$, while FONPPT gives a very good description  near the spacelike axis, which gradually deteriorates for points closer to the timelike axis, {\em i.e.} $\theta=0$. As explained in \cite{CaFi2009,CaFi2011}, this behaviour is due to the large imaginary parts of the logarithms in the coefficients  $c_{n, {\rm FO}}^{(jk)}(s)$, which follow from (\ref{eq:fopt}) and  (\ref{BFO}).  For the imaginary part of the Adler function  the results are similar \cite{CaFi2009,CaFi2011}.

\begin{table}[!htp]\label{table:tab1}
\caption{The difference $\delta^{(0)}-\delta_{\rm exact}^{(0)}$  for the reference model \cite{BeJa,BeBoJa} calculated for
  $\alpha_s(M_\tau^2)=0.34$ with CI and RGS  non-power expansions  truncated at order $N$. $\delta_{\rm exact}^{(0)}=0.2371$.}\vspace{0.2cm}
{\begin{tabular}{@{}cllllllll@{}} \hline 
 $N$ & CI $w_{12}$ & RGS $w_{12}$ & CI $w_{13}$ & RGS $w_{13}$ & CI $w_{1\infty}$ & RGS $w_{1\infty}$ & CI $w_{23}$  & RGS  $w_{23}$ \\ \hline
2 &   -0.0394    &-0.0347 &    -0.0301    &-0.0239 &   -0.0488   &-0.0417 &      -0.0248  & -0.0177\\
3&  -0.0362     &-0.0333 &    -0.0341    &-0.0301&    -0.0396      &-0.0349&        -0.0343            & -0.0303\\ 
4&   -0.0108    &-0.0089&    -0.0177     & -0.0142&     -0.0083     &-0.0067&      -0.0165            & -0.0132 \\ 
5&    -0.0081   &-0.0070&    -0.0103     &-0.0086&    -0.0061       &-0.0058&       -0.0079             &-0.0070 \\ 
6&   -0.0047    &-0.0073 &  -0.0065      &-0.0071&     -0.0050       &-0.0064&        -0.0052             &-0.0072\\
7&  -0.0032     &-0.0059&   -0.0040      &-0.0057&      -0.0038      &-0.0056&        -0.0026   &-0.0044 \\ 
8&  -0.0032     &-0.0041&     -0.0028    &-0.0035&     -0.0030       &-0.0041&        -0.0024             & -0.0011 \\ 
9&   -0.0030    &-0.0023&   -0.0023      &-0.0019&        -0.0025    &-0.0028&        -0.0024    & -0.0010\\ 
10&   -0.0020   &0.0014  &    -0.0023     &-0.0012&        -0.0023   &-0.0020&       -0.0018  &   0.0004\\
11&  -0.0012   &0.0036 &    -0.0023      &-0.0008&       -0.0022    &-0.0016&      -0.0023                  & -0.0009\\
12&  -0.0009   &0.0031 &   -0.0020      &-0.0006&         -0.0022    &-0.0015&         0.0003  &  0.0005\\ 
13&  -0.0009   &0.0026 &   -0.0016       &-0.0004&       -0.0022    &-0.0015&    -0.0023      & -0.0005\\ 
14&  -0.0007   &0.0018 &   -0.0010       &-0.0003&       -0.0022          &-0.0015&      0.0024   &  -0.0011\\ 
15&  -0.0004   &0.0006 &  -0.0005        & -0.0002&     -0.0021  &-0.0015&     -0.0015     &  0.0044\\ 
16&  -0.0003   & 0.0001&   -0.0002       & $-7 \cdot 10^{-6}$&   -0.0020            & -0.0015&   -0.0028       &-0.0131\\
17&  -0.0003   & -0.0004&  0.0001      & $4 \cdot 10^{-6}$ & -0.0019   &-0.0014&     0.0162  &  0.0238\\ 
18&  -0.0003   &-0.0013&    0.0002     &-0.0001&                    -0.0017       & -0.0013&      -0.0445 &  -0.0310\\ 
\hline
\end{tabular}\label{tab:tab1}}\vspace{0.4cm}
\end{table} 

It turns out that RGSNPPT gives results similar to those of CINPPT  \cite{Abbas:2013}.  This is illustrated in Table \ref{tab:tab1}, which shows the good perturbative convergence of these expansions for the integral (\ref{eq:delta0}), for various conformal mappings defined in the previous section. 

\begin{figure}\begin{center}\includegraphics[width=6.cm]{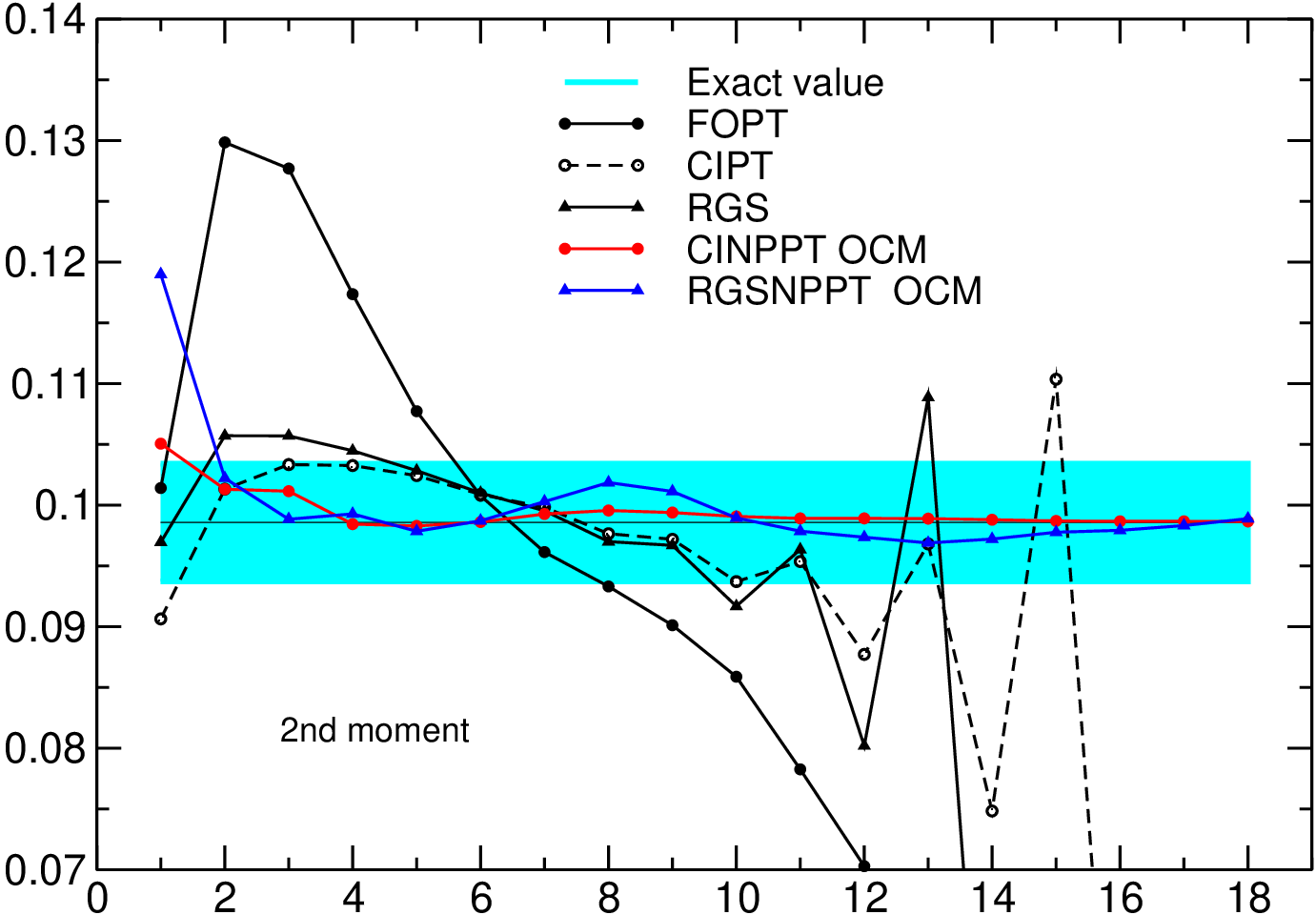}\hspace{1.5cm}\includegraphics[width=6.cm]{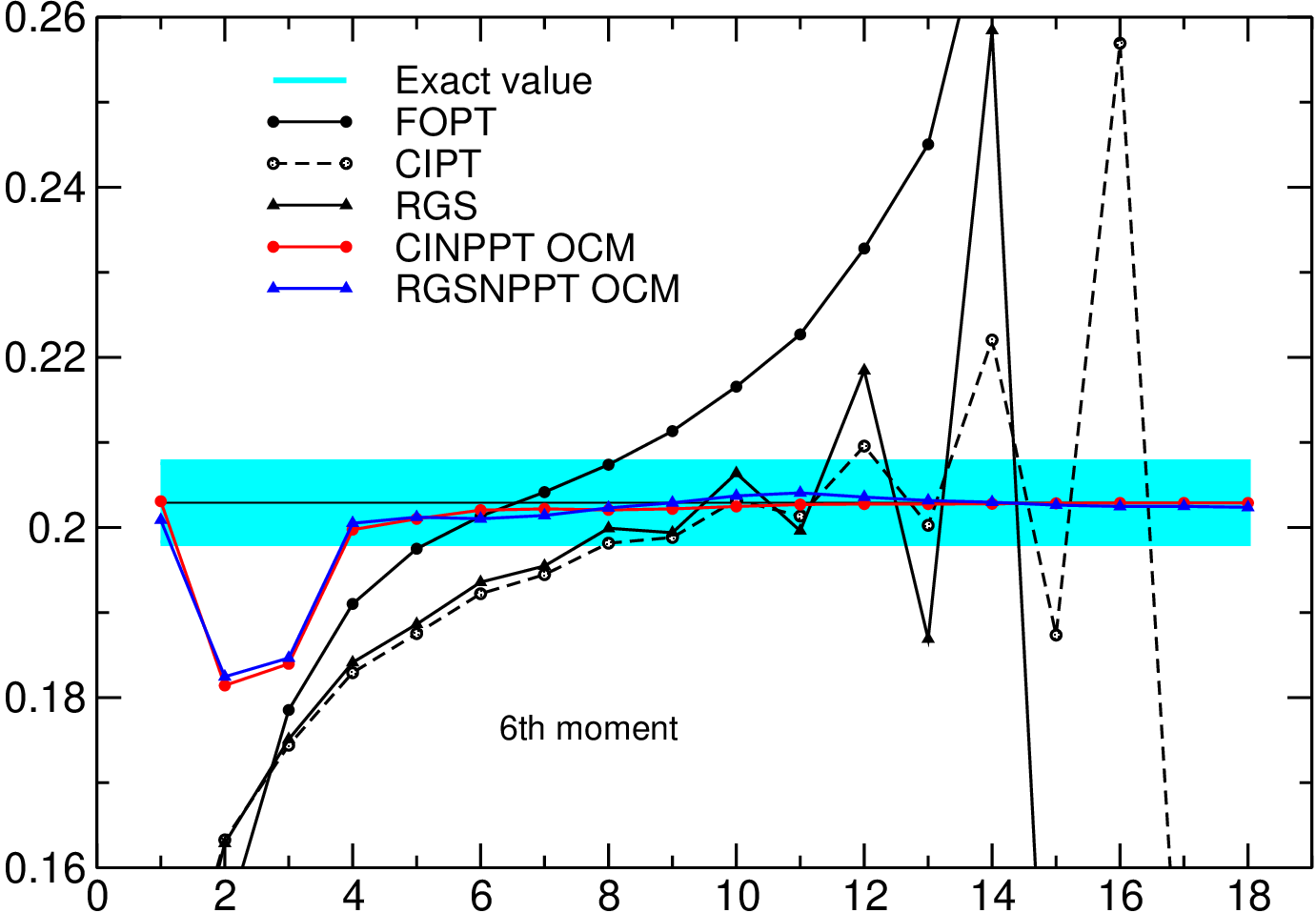}\caption{Perturbative expansions of the  moment $M_2$ (left) and  $M_6$ (right), in the notation of \cite{BeBoJa}. }\label{fig:4}\end{center}\end{figure}

\begin{figure}\begin{center}\includegraphics[width=6.cm]{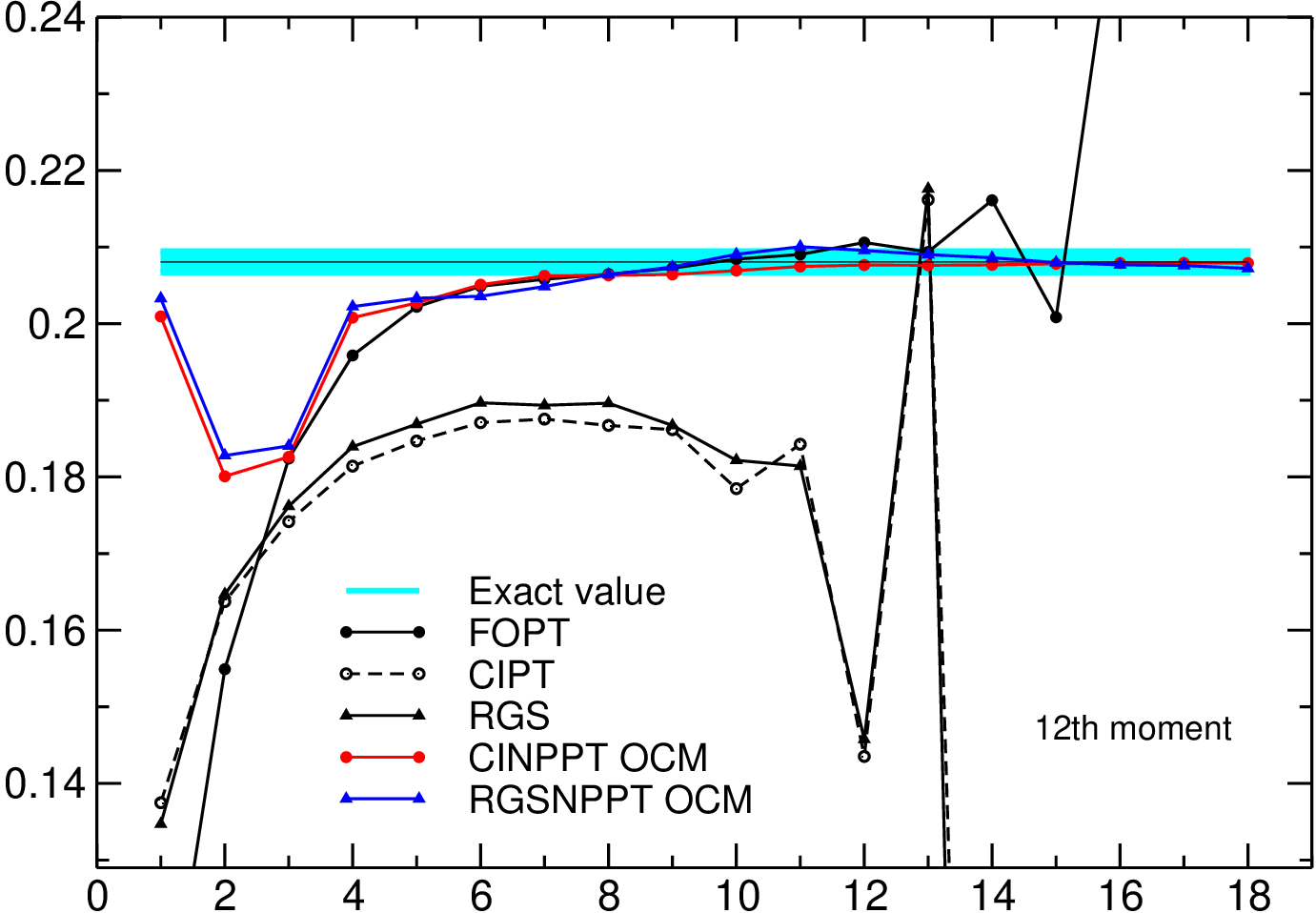} \hspace{1.5cm}\includegraphics[width=6.cm]{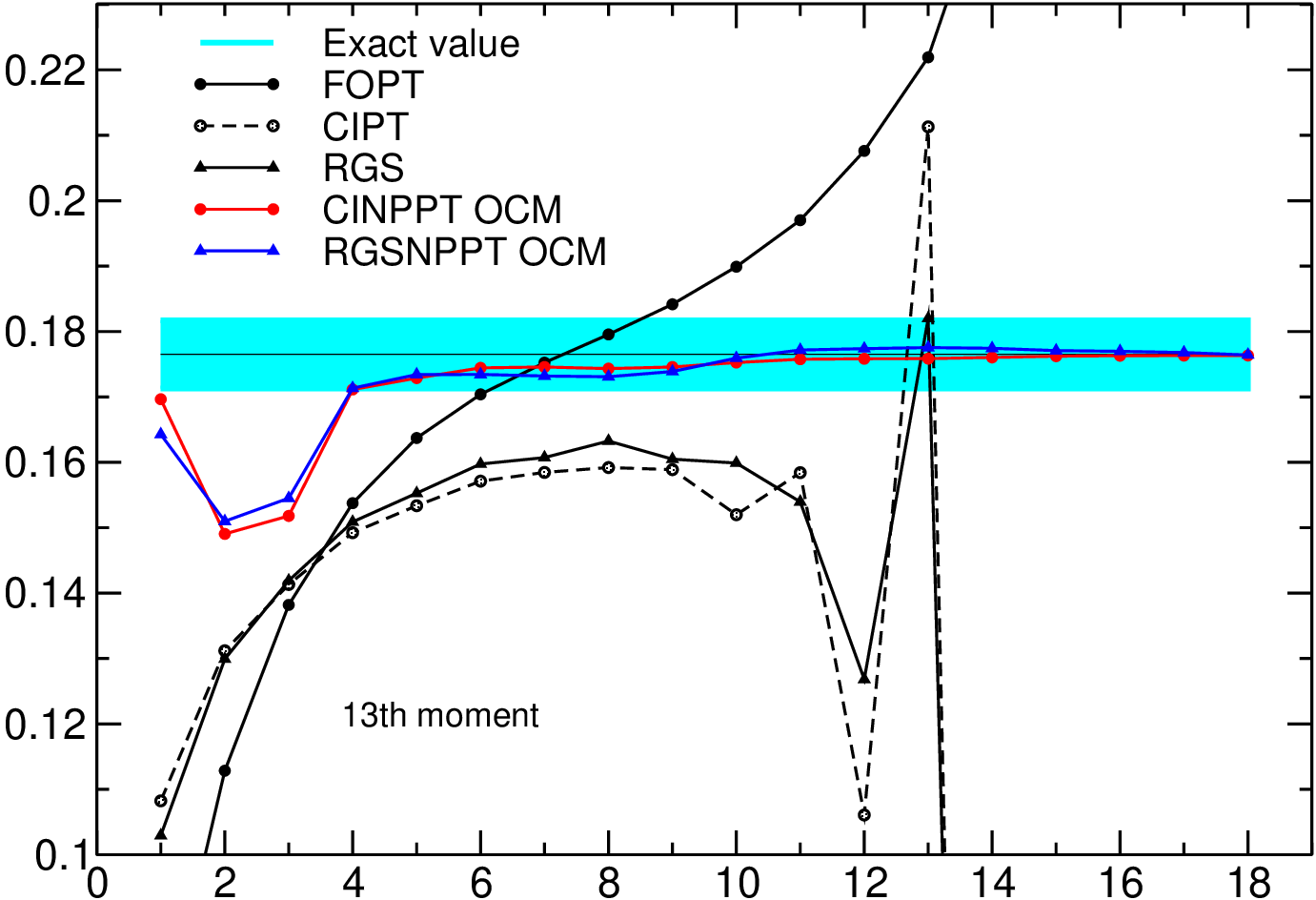}\caption{Perturbative expansions of the  moment $M_{12}$ (left) and  $M_{13}$ (right). }\label{fig:5}\end{center}\end{figure}

I end this section with a short discussion of the moments of the $\tau$ spectral function. Figs. \ref{fig:4} and \ref{fig:5}   show several moments considered recently in \cite{BeBoJa}, calculated with  various expansions. To facilitate the comparison with \cite{BeBoJa}, I took $\alpha_s(M_\tau^2)=0.3186$ in this calculation. The curves prove in an impressive way the good approximation achieved with the CI and RGS non-power expansions based on the optimal mapping (OCM), even for moments for which the standard CI, FO and RGS expansions fail badly. A detailed analysis will be reported elsewhere \cite{Abbas:mom}.

\section{Determination of $\alpha_s(M_\tau^2$)}\label{sec:alphas}
As shown in the previous section, the contour-improved and the renormalization-group summed non-power expansions (\ref{eq:cinppt}) and  (\ref{eq:rgsnppt}) have very good convergence properties, providing the best theoretical frame for the extraction of  the strong coupling from the $\tau$ hadronic width. The determination presented below relies on the known coefficients $c_{n,1}$ given in (\ref{eq:cn1}), the conservative choice  $c_{5,1}=283\pm 283$ for the next coefficient \cite{BeJa,Beneke_Muenchen}, and the phenomenological input \cite{Beneke_Muenchen}  
\begin{equation}\label{eq:delph}
 \delta^{(0)}_{\rm phen}=0.2037\pm 0.0040_{\rm exp}\pm 0.0037_{\rm PC}.
\end{equation}
   As reported in \cite{CaFi2011},  the CINPPT expansions (\ref{eq:cinppt}), with four choices of the conformal mappings, namely  $\wt w_{12}(u)$, $\wt w_{13}(u)$,  $\wt w_{1\infty}(u)$ and $\wt w_{23}(u)$, lead to the average 
\beq\label{eq:cifinal}\alpha_s(M_\tau^2)= 0.3195~^{+0.0189}_{- 0.0138}\,,\eeq
where the errors due to experiment, the PC corrections, the uncertainty of the coefficient $c_{5,1}$, the higher perturbative coefficients of the $\beta$-function and the choice of the scale were combined in quadrature. 
Similarly,  the average of the values obtained with  the RGSNPPT expansions (\ref{eq:rgsnppt}) reads \cite{Abbas:2013} 
\beq\label{eq:rgsfinal} \alpha_s(M_\tau^2)= 0.3189~^{+ 0.0145}_{-0.0115 }. \eeq
 In both cases the main part of the error is produced by the uncertainty of $c_{5,1}$ (assumed, as in \cite{BeJa,Beneke_Muenchen}, to be quite large),  while the variation of the scale and the inclusion of a five-loop coefficient in the $\beta$-function expansion have very small effects.

Combining in a conservative way the values given in (\ref{eq:cifinal}) and (\ref{eq:rgsfinal})  we obtain our final prediction: 
\beq\label{eq:final}\alpha_s(M_\tau^2)= 0.3192~^{+ 0.0167}_{-0.0126},~~\Rightarrow~~ \alpha_s(M_Z^2
)= 0.1184~^{+0.0020}_{-0.0016}\,.\eeq
The central value of $\alpha_s(M_Z^2)$ obtained by evolving our prediction for $\alpha_s(M_\tau^2)$  concides with the new world average  $\alpha_s(M_Z^2)=0.1184\pm 0.0007$ quoted in \cite{RPP}, while the errors of (\ref{eq:final}) are larger by factors between two and three.
\section{Summary and conclusions}
The $\tau$ hadronic width is a good observable for extracting the strong coupling, since it  receives small contributions from the nonperturbative power corrections and the possible breakdown of the OPE near the timelike axis. Starting from the two remaining ambiguities of perturbative QCD, RG prescription and the divergence of the series, I brought arguments in favour of a new, non-power perturbative expansion,  which simultaneously  implements renormalization-group summation and  the known location and nature of the first singularities of the expanded function in the Borel plane.   Numerical studies on models show that the countour-improved and the renormalization-group summed non-power expansions, denoted here as CINPPT and RGSNPPT, describe very well the Adler function and the moments of the spectral function, including some that are poorly described by the standard expansions. Therefore, 
these expansions provide a solid theoretical frame for the  determination of $\alpha_s(M_\tau^2)$ from the $\tau$ hadronic width.  Our prediction is given in (\ref{eq:final}). The error is larger that that obtained from $\tau$ decays with the standard expansions, but reflects better the uncertainty related to the series truncation. The calculation or a more precise estimate of the five-loop coefficient $c_{5,1}$ is crucial for reducing the  error.

\section*{Acknowledgments}
 I thank Prof. K.K. Phua and the workshop organizers for the kind hospitality at the
Institute for Advanced Studies, Nanyang Technological University, Singapore.
This talk is based on research done in  collaborations with Jan Fischer, Gauhar Abbas and B. Ananthanarayan. The work was supported by the Ministry of National Education under Contracts PN 09370102/2009 and Idei-PCE No 121/2011.

\end{document}